\documentclass[10pt,journal,onecolumn]{IEEEtran}

\usepackage{graphicx, bm, amsmath, siunitx, longtable,tabularx, float, mathtools, booktabs, eqnarray, color, amsfonts, amssymb, graphics, setspace, makecell}

\usepackage{cite}  

\usepackage[version=4]{mhchem}
\pdfoutput=1

\newcommand{\B}[1]{{\bm #1}}
\newcommand{\T}{^{\mbox{\tiny T}}}
\newcommand{\Ts}{^{\,\mbox{\tiny T}}}

\DeclareMathOperator{\diag}{diag}
\DeclareMathOperator{\Span}{span}
\DeclareMathOperator{\atan2}{atan2}

\usepackage{etoolbox}

\begin{document}

\title{Recursive Star-Identification Algorithm using an Adaptive SVD-based Angular Velocity Estimator$^*$\thanks{$^*$This work was paritally supported by a NASA Space Technology Research Fellowship, Johnston [NSTRF 2019] Grant \#: 80NSSC19K1149, Leake [NSTRF 2019] Grant \#: 80NSSC19K1152, and the NASA-MSFC by the Award \#: 80NSSC19P1369}}

\author{Hunter Johnston$^{1}$\thanks{$^{1}$ Aerospace Engineering, Texas A\&M University, College Station, TX, 77843 USA (hunterjohnston@tamu.edu, leakec@tamu.edu, mortari@tamu.edu)}, \and Carl Leake$^{1}$, \and Marcelino M. de Almeida$^{2}$\thanks{$^{2}$ Aerospace Engineering and Engineering Mechanics, University of Texas at Austin, Austin, Texas 78712, USA (marcelino.malmeidan@utexas.edu)}, \and Daniele Mortari$^{1}$}

\maketitle

\begin{abstract}
    This paper describes an algorithm obtained by merging a recursive star identification algorithm with a recently developed adaptive SVD-based estimator of the angular velocity vector (QuateRA). In a recursive algorithm, the more accurate the angular velocity estimate, the quicker and more robust to noise the resultant recursive algorithm is. Hence, combining these two techniques produces an algorithm capable of handling a variety of dynamics scenarios. The speed and robustness of the algorithm are highlighted in a selection of simulated scenarios. First, a speed comparison is made with the state-of-the-art lost-in-space star identification algorithm, Pyramid. This test shows that in the best case the algorithm is on average an order of magnitude faster than Pyramid. Next, the recursive algorithm is validated for a variety of dynamic cases including a ground-based ``Stellar Compass'' scenario, a satellite in geosynchronous orbit, a satellite during a re-orientation maneuver, and a satellite undergoing non-pure-spin dynamics.
\end{abstract}


\section{Introduction}

One of the classic scenarios for space-operating star trackers is the recursive scenario. In the recursive scenario, the attitude dynamic of a spacecraft is associated with a small  angular velocity, and the star identification (Star-ID) was completed at some previous time such that the stars previously identified are close to where they are currently observed. The information from this previous Star-ID, namely the locations of the stars at the previous time and their IDs, coupled with an estimate of the angular velocity can be used to predict where the stars will appear in the frame at the current time. This information can be used to identify some or all of the stars in the current frame. Moreover, the recursive procedure for identifying a star takes less time and requires less computation than identifying a star in the lost-in-space scenario.

The lost-in-space scenario is another classic star tracker operation mode wherein there is insufficient or non-existent information regarding the locations or IDs of stars in the frame. This scenario may be encountered anytime the recursive Star-ID (RSI) cannot be used. For example, during the first attitude estimate made by the star tracker, if a sufficient amount of time has passed since the previous Star-ID such that the previous Star-ID information is rendered useless, or the angular velocity estimate is inaccurate or non-existent. The attentive reader will notice that a lost-in-space algorithm (LISA) must be paired with the recursive algorithm to identify the stars initially and to act as a fail-safe whenever the recursive algorithm cannot perform the Star-ID. The Pyramid algorithm \cite{Pyramid} is the state-of-the-art LISA in terms of speed and robustness to noise (e.g. spikes in the camera frame that appear as fake stars), and is summarized in Appendix \ref{app:Pyramid}.

As previously mentioned, the recursive algorithm requires an accurate estimation of the angular velocity vector. Usually, this estimate comes from an external source, such as a gyroscope or Kalman filter. Spacecrafts are typically instrumented with one or multiple gyroscopes, but these come with a lifetime expectancy and eventually fail (for example, the Hubble Space Telescope currently has three remaining operational gyroscopes from the six that were installed in 2009 \cite{garner_2018}). On the other hand, Kalman filters can still be used to estimate the angular rate of a spacecraft, but they may not converge (or even diverge) when the dynamics do not fall within the nominal working range. 

Assuming that gyroscopic measurement are not available, Kalman filter designs perform particularly poorly if the inertia properties of the spacecraft are unknown, which might happen due to fuel consumption, changes in position of hardware, payload deployments or payload capture. Many of the existing angular velocity estimators \cite{salcudean1991globally,oshman2003spacecraft} rely on the knowledge of the spacecraft's inertia properties and torque parameters. An exception can be made for the \textit{derivative approach} proposed by Ref.~\cite{bar2001classification}, but as the author recognizes, the estimator can produce considerable error due to the presence of measurement noise. Ref.~\cite{bar2007rigid} presents the Pseudolinear Kalman Filter (PSELIKA), which is an estimation algorithm that does not depend on knowledge of the inertia matrix or input torques. However, PSELIKA is proposed with the goal of ``simplicity rather than accuracy'' \cite{bar2007rigid}, serving as a relatively coarse angular velocity estimator for control loop damping purposes.

An alternative solution for gyroless angular velocity estimation is to use methods based on the Multiplicative Extended Kalman Filter (MEKF) \cite{lefferts1982kalman, gai1985star, markley2002attitude}, since they do not necessarily rely on the inertia properties of the spacecraft (they are kinematic-based). Still, these methods require proper initialization and tuning for ranges of accepted angular accelerations, and one might want to perform a backward smoothing process \cite{psiaki2005backward} for proper convergence.

The work in Ref.~\cite{psiaki2009generalized} generalizes Wahba's problem by proposing a new problem that accepts sequential vector measurements instead of the traditional simultaneous ones (see Ref.~\cite{markley2000quaternion} and the references therein). These generalizations imply the need to estimate the initial orientation and angular velocity (not only orientation, as in Wahba's problem) based on multiple sequential vector measurements. The following problems are proposed in Ref.~\cite{psiaki2009generalized}:
\begin{itemize}
    \item First Generalized Wahba's Problem (FGWP) - The system is in pure spin with a known spin-axis but unknown spin rate. Ref.~\cite{psiaki2009generalized} presents a closed-form solution to this problem based on two measurements. Furthermore, Ref.~\cite{saunderson2015convex} uses semidefinite programming to solve FGWP for more than two measurements.
    \item Second Generalized Wahba's Problem (SGWP) - The system is tumbling (torque-free) with a known inertia matrix. This system is proved to be observable with at least three vector measurements, but no solution is provided within Ref.~\cite{psiaki2009generalized}. A solution to the three-vector measurement problem is provided in Ref.~\cite{hinks2011solution}, and a numerical solution is provided in Ref.~\cite{psiaki2012numerical} for four or more measurements.
\end{itemize}

The current work diverges from both FGWP and SGWP in that we do not assume knowledge of the spin-axis (as in FGWP) nor of the inertia matrix (as in SGWP). In essence, the problem that we propose and solve is similar to SGWP without the requirement of knowledge of the system's inertia matrix. Recently, a novel SVD-based angular velocity estimation filter, QuateRA \cite{Koh, Akella, NonCoop, QuateRA}, was developed, enabling an adaptive estimation of the angular velocity vector over a wider range of attitude dynamics without requiring knowledge of the spacecraft's inertial properties.

The proposed process is the following: Pyramid performs a sequence of lost-in-space Star-IDs to estimate a sequence of attitudes. These attitude estimations are then used by QuateRA to estimate the angular velocity vector. Then, the recursive Star-ID algorithm (RSI), using the angular velocity estimate from QuateRA and most recent Pyramid Star-ID, predicts the star directions at the current time, and then performs a Star-ID with the actual observed star directions. In the nominal scenario, the attitude estimation made by RSI is then used by QuateRA to update the angular velocity measurement. Afterwards, QuateRA and RSI can work in tandem to continually estimate the attitude and the angular velocity. Performing the Star-ID in this way is faster and less computationally intensive than a LISA such as Pyramid.

Therefore, the combination of QuateRA and RSI enables the estimation of a spacecraft's attitude and attitude rate with one sensor and during a wide range of attitude dynamics, such as during attitude maneuvers. Moreover, since QuateRA is an adaptive method, for each angular velocity estimate it can increase, maintain, or decrease the number of attitude measurements it uses. This is the adaptive feature of QuateRA, which keeps the maximum number of consecutive attitude measurements that approximately describe pure spin attitude dynamics. In the case of a failure, Pyramid is available to solve the lost-in-space problem and re-initialize the ``RSI + QuateRA'' process. 

This remainder of the article is structured as follows. First, the theory for the RSI algorithm is summarized based on Ref. \cite{Carl}. To highlight the robustness, this recursive algorithm is successfully tested during a full rotation period on torque-free rigid-body dynamics in the presence of random fake stars (spikes). Next, the QuateRA theory \cite{Koh, Akella, NonCoop, QuateRA} is summarized and implemented alongside the RSI algorithm. Afterwards, numerical tests are presented to highlight the speed and robustness of the algorithm. First, the RSI algorithm is compared to the Pyramid Star-ID algorithm to determine the maximum expected time saved when using the recursive technique. Next, four specific scenarios over an array of varying dynamics are presented to validate the method and show its range of applicability. 

\section{The Recursive Star-Identification Problem}

\begin{figure}[ht]
    \centering\includegraphics[width=0.7\linewidth]{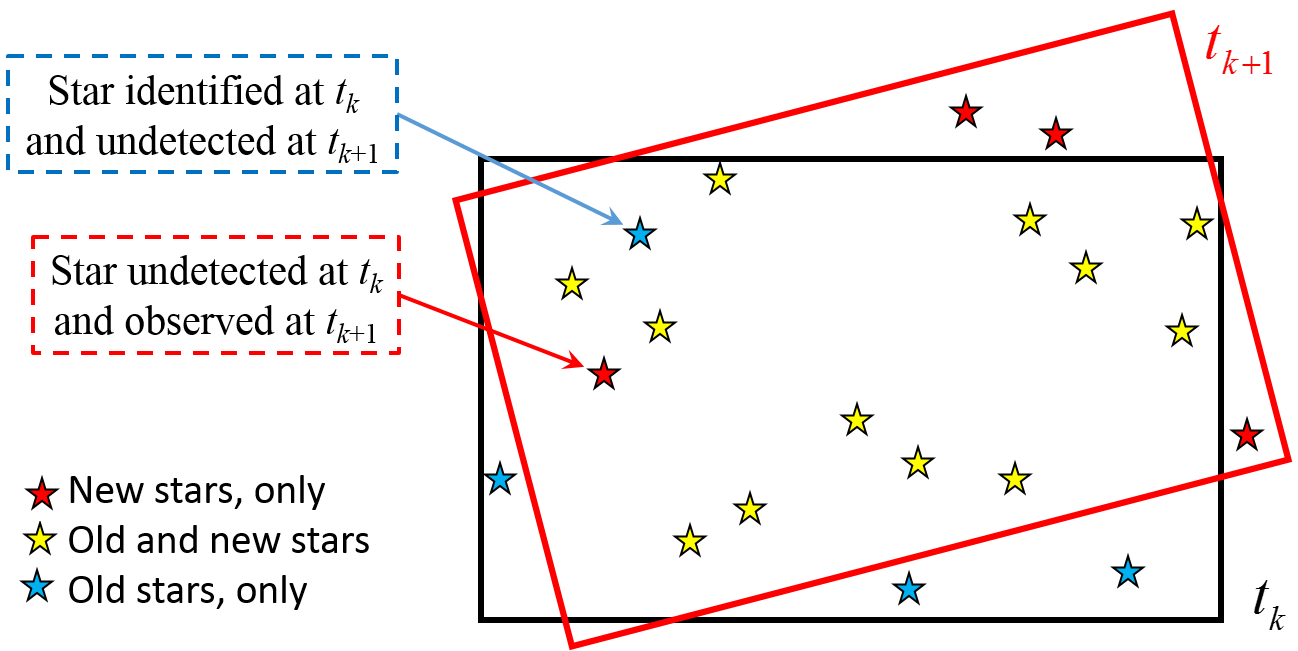} \\
    \caption{Recursive star-identification problem geometry}
    \label{Fig1}
\end{figure}

This section briefly summarizes the RSI algorithm. For a more thorough explanation and numerical validation see Ref. \cite{Carl}. Let $B_1 := \left[\B{b}_1^1, \B{b}_2^1, \cdots, \B{b}_{n_1}^1\right]$, be the matrix containing the $n_1$ observed star directions (unit-vectors) identified at time $t_1$, meaning $B_1 \approx C_1 \ R_1$, where $C_1$ is the Direction Cosine Matrix (DCM) that transforms vectors from the inertial frame to the camera reference frame, and $R_1 := [\B{r}_1, \B{r}_2, \cdots, \B{r}_{n_1}]$ is the matrix containing the inertial cataloged star directions (unit-vectors) at time $t_1$. Additionally, let $\B{\omega}_1$ be the angular velocity vector at time $t_1$, and let $B_2 := [\B{b}_1^2, \B{b}_2^2, \cdots, \B{b}_{n_2}^2]$ be the matrix containing the $n_2$ stars observed at time $t_2$. The geometry of the recursive problem for the two generic subsequent times, $t_k$ and $t_{k+1}$, is shown in Fig. \ref{Fig1}.

Under the assumption of small angular velocity direction variations, a recurrent star $\B{b}_i$ changes its angular position (from $t_k$ to $t_{k+1}$) by an angular deviation within the bounds,
\begin{equation*}
    \vartheta_{\min} < \vartheta_k \le \left(\Omega_k + 3\sigma_{\omega}\right) (t_{k+1} - t_k)
\end{equation*}
where $\vartheta_{\min}$ is the minimum inter-star angular separation, $\B{\omega}_k$ is the angular velocity at time $t_k$ and $\sigma_{\omega}$ represents one standard deviation of the angular velocity vector modulus. Note, in this paper we define the angular velocity vector such that = $\B{\omega} := \Omega \, \vec{\B{\omega}}$ where $\Omega$ is the modulus and $\vec{\B{\omega}}$ is the direction of the angular velocity vector. This notation is used because in future sections, the hat notation ($\hat{\text{\phantom{a}}}$) is used to signify a measurement. 

Now, let us consider the time variation, $t_{k+1} - t_k$, to be small enough so that the pure spin assumption (d$\vec{\B{\omega}}_k / {\rm d} t \approx \B{0}$) is a good approximation within the time interval. Under this assumption, the generic star $\B{b}_i$ observed at time $t_k$ is expected to be observed at time $t_{k+1}$ as,
\begin{equation*}
    E\{\B{b}_i\} = \delta C (\vec{\B{\omega}}_k, \varphi_k) \ \B{b}_i \qquad \text{where} \qquad \varphi_k = \Omega_k \, (t_{k+1} - t_k)
\end{equation*}
and where,
\begin{equation}\label{eq88}
    \delta C (\vec{\B{\omega}}_k, \varphi_k) = I_{3\times 3} \cos\varphi_k + (1 - \cos\varphi_k) \vec{\B{\omega}}_k \vec{\B{\omega}}_k\T - [\vec{\B{\omega}}_k\times] \sin\varphi_k,
\end{equation}
is the matrix performing a rigid rotation about the angular velocity direction, $\vec{\B{\omega}}_k$, through the rotation angle, $\varphi_k$. Therefore, the observed stars in the observed stars matrix $B_k = \left[\B{b}_1, \; \B{b}_2, \cdots, \B{b}_{n}\right]_k$ are expected to be seen at time $t_{k+1}$ at,
\begin{equation*}
    E \{ B_{k+1} \} = \delta C (\vec{\B{\omega}}_k, \varphi_k) \ B_k.
\end{equation*}
Once the expected star directions, $E \{ B_{k+1} \}$, are computed, then an algorithm, for example predictive centroiding \cite{PC}, matches the recurrent stars with the current true scenario, $B_{k+1}$. This algorithm, considers the new star, $\B{b}_j$, recurrent to the $E \{\B{b}_i \}$ star, if and only if,
\begin{equation}\label{eq06}
    \B{b}_j\T E\{ \B{b}_i \} > \cos\varepsilon \qquad \text{and} \qquad \B{b}_j\T E\{ \B{b}_\ell \} < \cos\varepsilon, \, \forall \, \ell\ne i,
\end{equation}
where $\varepsilon$ is an angular tolerance that depends on the accuracy of the estimated angular velocity vector $\B{\omega}_k$.

If at least three stars satisfy Eq. (\ref{eq06}), then the remaining unidentified stars will be identified (or discarded if they are spikes) the same way Pyramid identifies the remaining stars once the basic star triangle has been found. Otherwise, the recursive algorithm is aborted and the new scenario is given to Pyramid to solve for the lost-in-space case. After the identification of all stars, two distinct checks are performed to validate the identification:
\begin{enumerate}
    \item The new attitude estimated, $C_{k+1}$, is associated with a ``sufficiently'' small Wahba's cost function value, and
    \item The relative corrective attitude, $C_k \Ts C_{k+1}$, has a principal angle, $\Phi_k$, ``sufficiently'' close to $\varphi_k$,
    \begin{equation*}
        \hat{C}_{k+1} = \delta C (\vec{\B{\omega}}_k, \varphi_k) \ C_k \qquad \to \qquad \Phi_k = \cos\left[\dfrac{{\rm trace} (C_k \Ts C_{k+1}) - 1}{2}\right] \approx \varphi_k
    \end{equation*}
\end{enumerate}
where $\hat{C}_{k+1}$ is the estimated $C_{k+1}$ attitude. If both of these conditions are met, then the identification performed by the recursive algorithm is considered correct. Then, the $B_{k+1}$ star directions along with their associated inertial vectors $R_{k+1}$ can be used in a variety of available attitude estimation techniques; in this paper we use the q-method \cite{qMethod}. Once the attitude is determined, the information can be sent to a filter (i.e. Kalman filter, QuateRA, etc.) to improve the estimation accuracy, and then to the control system.

\begin{figure}[ht]
    \centering\includegraphics[width=\linewidth]{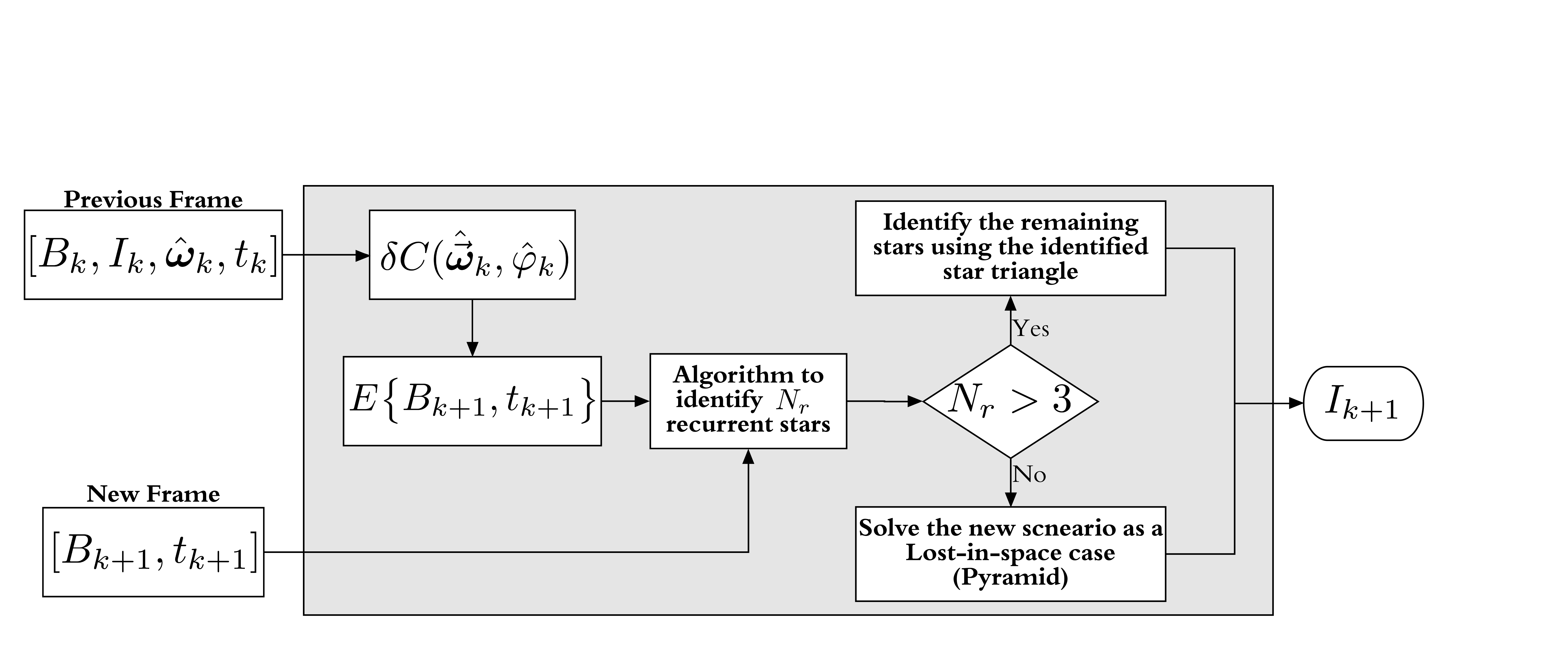}
    \caption{Recursive Star-Identification flowchart.}
    \label{fig:flow}
\end{figure}

A flowchart is provided in Figure \ref{fig:flow} to summarize the major steps in the RSI algorithm. The output of the RSI algorithm is the vector of indices $I_{k+1}$ associated with the $B_{k+1}$ stars, which are used in conjunction with the inertial vectors $R_{k+1}$ to compute the attitude.

The star catalog is made of all stars with a magnitude lower than the star tracker magnitude threshold, and the database needed to implement the $k$-vector \cite{original} range searching used by Pyramid \cite{Pyramid}. This database contains all the admissible star pairs, meaning observable, for the camera field-of-view and the $k$-vector value and line parameters. The recursive algorithm input contains variables associated with the previous identification. These are, specified at time $t_k$, the angular velocity vector $\B{\omega}_k$, all observed and identified stars, $B_k$, and the identification index vector, $I_k$. In addition, the observed unidentified stars, $B_{k+1}$, at the current time, $t_{k+1}$, are inputs to the recursive algorithm. Using the knowledge of $\vec{\B{\omega}}_k$ and $\varphi_k$, matrix $\delta C (\vec{\B{\omega}}_k, \varphi_k)$ as specified in Eq. (\ref{eq88}) can be computed. This matrix allows us to estimate where the previously observed and identified stars should be expected, $E \{B_{k+1}\}$. Because of the errors on the estimated attitude dynamics, some of these expected stars fall outside the sensor FOV. For this reason they are removed from $E \{B_{k+1}\}$. Using only those stars that fall within the sensor FOV, a simple algorithm validates which of the actual observed stars, $B_{k+1}$, match with those expected, $E \{B_{k+1}\}$. If more than 3 actual stars are validated by this algorithm, then the remaining observed stars are identified using the same logic that Pyramid uses to perform the identification once the $i$, $j$, $k$, and $r$ stars have been identified (step 3 of the Pyramid algorithm shown in Appendix \ref{app:Pyramid}). Otherwise, the process is aborted and the Pyramid algorithm is called to identify the observed stars as in a lost-in-space scenario.

\section{The SVD-based Adaptive Angular Velocity Estimation Filter}

The second part to the proposed algorithm relies on an adaptive algorithm that estimates the direction of the angular velocity using a sequence of quaternions introduced in Ref. \cite{Koh}. While other techniques exist to estimate spacecraft angular velocity, such as Ref. \cite{opt} which uses sequences of star-field images, these methods do not leverage the known dynamics of the problem. The main idea for this estimation is based on the fact that, if the angular velocity of a rigid body does not change direction (i.e. the body is in pure spin), then the quaternion dynamics on the 4-D unit sphere lie on a fixed plane (i.e. all the quaternions lie on a 2-dimensional circle embedded in 4-dimensional space). Stated a different way, if a rigid body is known to be in pure spin and the initial quaternion is given, then a bijective relationship exists between the angular velocity direction and the osculating quaternion plane. This relationship, introduced in Ref. \cite{Koh}, has been extended to the continuous case \cite{Akella}, to non-cooperative space objects \cite{NonCoop}, and the theory was completed in Ref. \cite{QuateRA} by developing an optimal filter. In Ref. \cite{QuateRA}, the osculating quaternion plane is identified by the quaternion itself, $\B{q}$, and its time derivative, $\dot{\B{q}}$. 

\subsection{Summary of the Quaternion Regression Algorithm}

This section presents the Quaternion Regression Algorithm (QuateRA), as well as some crucial aspects used in its derivation in Ref. \cite{QuateRA}. QuateRA is an algorithm that estimates the angular velocity of a body in pure spin by using a sequence of measured quaternions. QuateRA first estimates the axis of rotation (AOR) $\hat{\vec{\B{\omega}}}$, then it uses the estimated AOR to estimate the angular velocity magnitude (AVM) $\hat{\Omega}$. Finally, the estimated angular velocity is calculated using $\hat{\B{\omega}} = \hat{\Omega} \, \hat{\vec{\B{\omega}}}$.

In order to estimate the AOR, QuateRA uses a geometric interpretation based on the solution to the quaternion kinematic equation for constant $\B{\omega}$,
\begin{align} \label{eq_avast:solution_qke_constant_omega}
	\B{q}(t) = \begin{bmatrix} I_{4\times 4} \, \cos\left(\dfrac{\Omega\delta t}{2}\right)  +  \sin\left(\dfrac{\Omega\delta t}{2}\right) [\vec{\B{\omega}_q} \otimes]\end{bmatrix}  \B{q}_0, \qquad {\rm where} \qquad \vec{\B{\omega}}_q = \begin{Bmatrix} \vec{\B{\omega}} \\ 0 \end{Bmatrix},
\end{align}
and where $\delta t \triangleq t - t_0$. Defining the vectors $\B{u}_1 \in \mathbb{S}^3 = \B{q}_0$ and $\B{u}_2 \in \mathbb{S}^3 = [\vec{\B{\omega}}_q \otimes] \B{q}_0$, we have that $\B{u}_1\T \B{u}_2 = \B{q}_0\T [\vec{\B{\omega}}_q \otimes] \B{q}_0$, where $\mathbb{S}^3$ indicates the unit-sphere in 4D space. Since $[\vec{\B{\omega}}_q \otimes]$ is a $4\times 4$ skew-symmetric matrix then $\B{u}_1\T \B{u}_2 = 0$, i.e., $\B{u}_1 \perp \B{u}_2$. Clearly, any $\B{q}(t)$ described by Eq. (\ref{eq_avast:solution_qke_constant_omega}) is a linear combination of $\B{u}_1$ and $\B{u}_2$, for all $t \in \mathbb{R}$. Hence, if we define the 4D hyperplane $\mathbb{P}(\B{u}_1,\B{u}_2) = \Span \{\B{u}_1, \B{u}_2\}$, then $\B{q}(t) \in \mathbb{P}(\B{u}_1,\B{u}_2), \, \forall \, t \in \mathbb{R}$. In addition, there exists a perpendicular plane $\mathbb{P}(\B{u}_3,\B{u}_4) = \Span \{\B{u}_3, \B{u}_4\}$, with $\B{u}_3, \B{u}_4 \in \mathbb{S}^3$ such that $\B{u}_4 = [\vec{\B{\omega}}_q \otimes] \B{u}_3$, where $\B{u}_3\T\B{q}(t) = \B{u}_4\T\B{q}(t) = 0,\, \forall \, t \in \mathbb{R}$.

Therefore, given a sequence of $n$ quaternion measurements $\bar{\B{q}}_i$, with $n \in \mathbb{N}_{\geq 2}$, QuateRA estimates the AOR by finding the optimal hyperplane that minimizes the distance to the measured quaternions. At a given time $t_k$, QuateRA constructs the measurement matrix with $n$ measurements $\bar{Q}_{k,n}$ as,
\begin{gather} \label{eq_avast:meas_matrix}
    \bar{Q}_{k,n} \triangleq \begin{bmatrix} \bar{\B{q}}_{k-n+1}, & \bar{\B{q}}_{k-n+2}, & \cdots, & \bar{\B{q}}_k\end{bmatrix}.
\end{gather}
where the number of quaternion measurements, $n$, used can be adapted during implementation based on the singular values of the Singular Value Decomposition (SVD) taken in the following steps. Note that the quaternions in each column of $\hat{Q}_{k,n}$ should belong to the estimated plane of rotation: $\hat{\B{q}}_i \in \mathbb{P} (\hat{\B{u}}_1, \hat{\B{u}}_2), \, i \in \{k-n+1, \cdots, k\}$. The quaternions $\hat{\B{q}}_i$ are estimated to minimize the total least squares cost function,
\begin{gather*}
    J_0 = \frac{1}{2}\begin{Vmatrix} \bar{Q}_{k,n} - \hat{Q}_{k,n} \end{Vmatrix}_F^2,
\end{gather*}
subject to $\hat{\B{q}}_i \in \mathbb{P}(\hat{\B{u}}_1,\hat{\B{u}}_2), \, \forall \, i \in \{k-n+1, \cdots, k\}$, where $\hat{\B{u}}_1$ and $\hat{\B{u}}_2$ define the optimally estimated plane of rotation. Assuming small angle approximation for the noise-polluted quaternion, Ref. \cite{QuateRA} shows that the optimization problem above is approximately equivalent to finding the unit-norm vectors $\hat{\B{u}}_1 \in \mathbb{S}^3$, $\hat{\B{u}}_2 \in \mathbb{S}^3$ such that $\hat{\B{u}}_1\T \hat{\B{u}}_2 = 0$, that maximize the following cost function,
\begin{align}\label{eq_avast:final_cost}
    J = \sum_{i = 1}^n \begin{bmatrix} \begin{pmatrix} \bar{\B{q}}_i\T \hat{\B{u}}_1\end{pmatrix}^2 + \begin{pmatrix} \bar{\B{q}}_i\T \hat{\B{u}}_2 \end{pmatrix}^2 \end{bmatrix} = \hat{\B{u}}_1\T \bar{Z} \hat{\B{u}}_1 + \hat{\B{u}}_2\T \bar{Z} \hat{\B{u}}_2,
\end{align}
where $\bar{Z} \triangleq \bar{Q}_{k,n}\bar{Q}\T_{k,n}$. Given $\hat{\B{u}}_1$, $\hat{\B{u}}_2$, the optimally estimated quaternions within $\hat{Q}$ are given by:
\begin{gather} \label{eq_avast:optimal_quaternions_plane}
    \hat{\B{q}}_i = \frac{1}{\sqrt{\begin{pmatrix} \bar{\B{q}}_i\T\hat{\B{u}}_1 \end{pmatrix}^2 + \begin{pmatrix} \bar{\B{q}}_i\T\hat{\B{u}}_2 \end{pmatrix}^2}}
    \begin{bmatrix} \begin{pmatrix} \bar{\B{q}}_i\T\hat{\B{u}}_1 \end{pmatrix}\hat{\B{u}}_1 +
                    \begin{pmatrix} \bar{\B{q}}_i\T\hat{\B{u}}_2 \end{pmatrix}\hat{\B{u}}_2\end{bmatrix}.
\end{gather}
Ref. \cite{QuateRA} proves non-uniqueness of the solution $\hat{\B{u}}_1$, $\hat{\B{u}}_2$ that maximizes Eq. (\ref{eq_avast:final_cost}). This holds because the solution can also be described by any other pair of vectors $\hat{\B{v}}_1 \in \mathbb{S}^3$, $\hat{\B{v}}_2 \in \mathbb{S}^3$ that satisfy $\hat{\B{v}}_1\T \hat{\B{v}}_2 = 0$ and $\hat{\B{v}}_1, \hat{\B{v}}_2 \in \mathbb{P}(\hat{\B{u}}_1,\hat{\B{u}}_2)$. A particular solution to the plane-fitting problem can be obtained through SVD of $\bar{Z} = \hat{U} \, \hat{\Sigma} \, \hat{U}\T$, where $\hat{U} \in \mathbb{R}^{4\times 4} = \begin{bmatrix} \hat{\B{u}}_1, & \hat{\B{u}}_2, & \hat{\B{u}}_3, & \hat{\B{u}}_4\end{bmatrix}$ contains the \textit{singular vectors of} $\bar{Z}$, and $\hat{\Sigma} = \diag \begin{pmatrix} \hat{\sigma}_1, \hat{\sigma}_2, \hat{\sigma}_3, \hat{\sigma}_4 \end{pmatrix}$ contains the \textit{singular values} of $\bar{Z}$, wherein $\hat{\sigma}_1 \geq \hat{\sigma}_2 \geq \hat{\sigma}_3 \geq \hat{\sigma}_4 \geq 0$. If $\hat{\sigma}_2 > \hat{\sigma}_3$, then $\hat{\B{u}}_1$ and $\hat{\B{u}}_2$ compose a solution to the  optimization problem in Eq. (\ref{eq_avast:final_cost}) and the optimal cost is given by $J^*(\hat{\B{u}}_1, \hat{\B{u}}_2) = \hat{\sigma}_1 + \hat{\sigma}_2$, with $\hat{\sigma}_1 = \hat{\B{u}}_1\T \bar{Z} \hat{\B{u}}_1$ and $\hat{\sigma}_2 =  \hat{\B{u}}_2\T \bar{Z} \hat{\B{u}}_2$. It is also true that $\hat{\sigma}_3 =  \hat{\B{u}}_3\T \bar{Z} \hat{\B{u}}_3$ and $\hat{\sigma}_4 = \hat{\B{u}}_4\T \bar{Z} \hat{\B{u}}_4$.

Having calculated the optimal hyperplane estimate $\hat{\mathbb{P}}(\hat{\B{u}}_1,\hat{\B{u}}_2)$, the optimal estimate for the AOR is given by,
\begin{gather} \label{eq_avast:optimal_aor}
    \hat{\vec{\B{\omega}}} = [\hat{\B{u}}_2 \otimes] \hat{\B{u}}_1^{-1}.
\end{gather}
The optimal quaternion estimates $\hat{\B{q}}_i \in \mathbb{P}(\hat{\B{u}}_1,\hat{\B{u}}_2), \, i \in \{k-n+1, \cdots, k\}$ can be re-parameterized as just an angle on the plane $\mathbb{P}(\hat{\B{u}}_1,\hat{\B{u}}_2)$. Taking $\hat{\B{u}}_1$ as a reference vector, the angle $\hat{\Phi}_i$ of any quaternion $\hat{\B{q}}_i$ with respect to $\hat{\B{u}}_1$ is given by:
\begin{gather} \label{eq_avast:phi_hat}
    \hat{\Phi}_i = 2 \cdot \atan2 \begin{pmatrix}{\hat{\B{q}}}_i\T \hat{\B{u}}_2, \,\,\, {\hat{\B{q}}}_i\T \hat{\B{u}}_1 \end{pmatrix}, \quad\quad i \in \{k-n+1, \cdots, k\}.
\end{gather}
Then, assuming the model,
\begin{gather*}
    \Phi_i = \Phi_0 + \Omega t_i = \begin{Bmatrix} 1, & t_i \end{Bmatrix} \begin{Bmatrix} \Phi_0 \\ \Omega\end{Bmatrix},
\end{gather*}
we can perform the least squares estimation, which leads to an estimate of the AVM $\hat{\Omega}$:
\begin{gather} \label{eq_avast:ls_phi0_omega}
    \hat{\B{X}} \triangleq \begin{Bmatrix} \hat{\Phi}_0 \\ \hat{\Omega}\end{Bmatrix} = \begin{bmatrix} H\T H\end{bmatrix}^{-1} H\T \hat{\B{\Phi}},
\end{gather}
where,
\begin{gather} \label{eq_avast:ls_phi0_omega2}
    H \triangleq \begin{bmatrix} 1 & \cdots & 1 \\ t_{k-n+1} & \cdots & t_k\end{bmatrix}\T, \qquad \text{and} \qquad \hat{\B{\Phi}} \triangleq \begin{Bmatrix} \hat{\Phi}_{k-n+1}, & \cdots, & \hat{\Phi}_k\end{Bmatrix}\T.
\end{gather}
In summary, QuateRA follows the algorithm below:
\begin{enumerate}
    \item Construct the measurement matrix $\bar{Q}_{k,n}$ as in Eq. (\ref{eq_avast:meas_matrix}) and calculate $\bar{Z} = \bar{Q}_{k,n}\bar{Q}\T_{k,n}$.
    \item Compute the SVD $\bar{Z} = \hat{U} \, \hat{\Sigma} \, \hat{U}\T$. The plane of rotation is defined by the first two columns of $\hat{U} = \begin{bmatrix} \hat{\B{u}}_1, & \hat{\B{u}}_2, & \hat{\B{u}}_3, & \hat{\B{u}}_4 \end{bmatrix}$.
    \item The optimal axis of rotation is defined as in Eq. (\ref{eq_avast:optimal_aor}): $\hat{\B{\omega}} = [\hat{\B{u}}_2 \otimes] \, \hat{\B{u}}_1^{-1}$.
    \item Compute the optimally estimated quaternions $\hat{\B{q}}_i, i \in \{1, \cdots, n\}$ on the plane $\hat{\mathbb{P}}(\hat{\B{u}}_1,\hat{\B{u}}_2)$ using Eq. (\ref{eq_avast:optimal_quaternions_plane}).
    \item For each quaternion $\hat{\B{q}}_i$ on the plane $\hat{\mathbb{P}}(\hat{\B{u}}_1,\hat{\B{u}}_2)$, compute the quaternion angle within the plane $\hat{\Phi}_i$ using Eq. (\ref{eq_avast:phi_hat}).
    \item Estimate the angular velocity $\hat{\Omega}$ and its associated covariance using Eqs. (\ref{eq_avast:ls_phi0_omega}) and Eq. (\ref{eq_avast:ls_phi0_omega2}). \emph{Note that the angles $\bar{\Phi}$ need to be unwrapped before performing the least squares estimation.}
\end{enumerate}

\section{Recursive Star-ID Algorithm with QuateRA}
\begin{figure}[ht]
    \centering\includegraphics[width=0.95\linewidth]{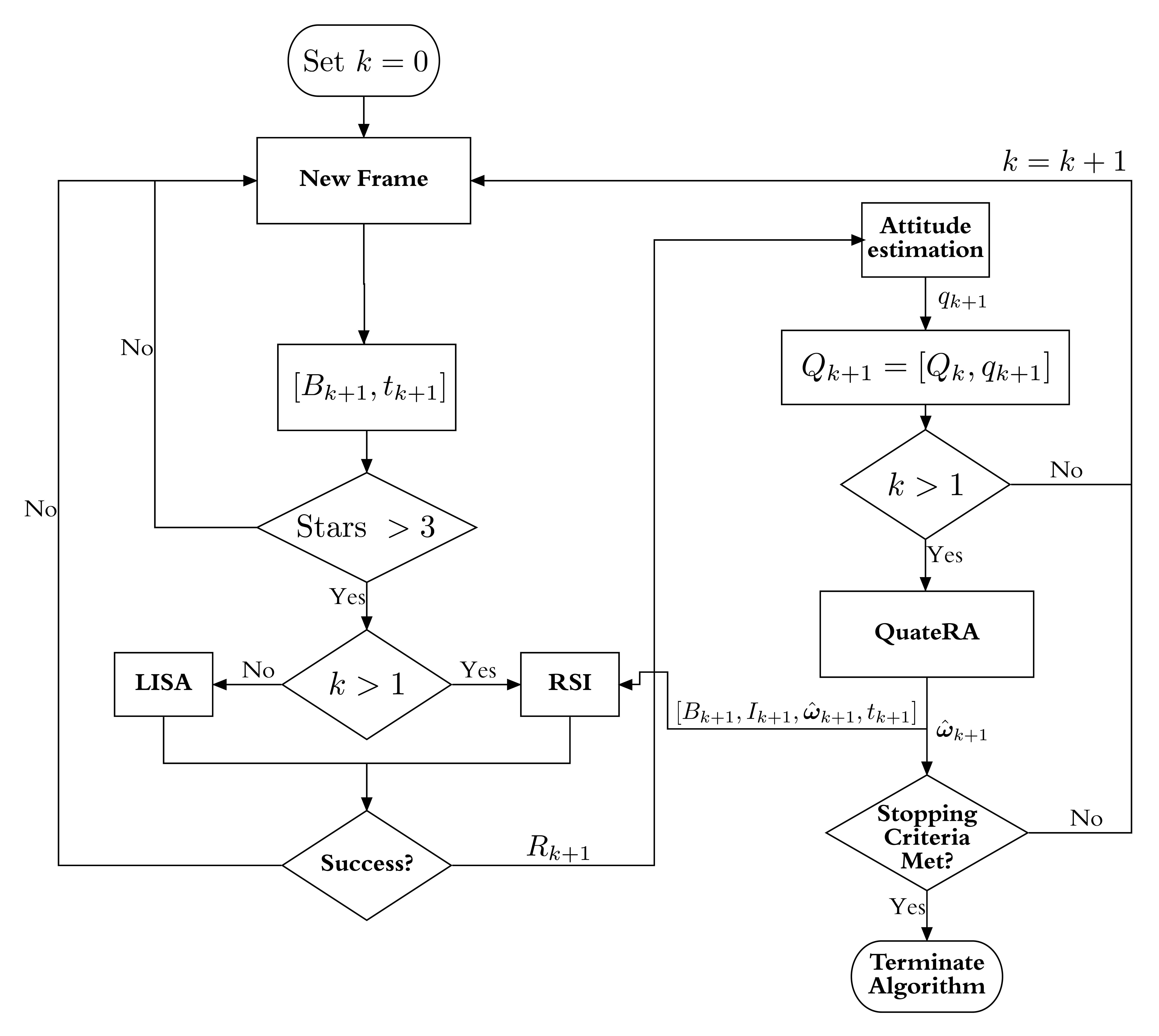}
    \caption{Full flow chart of the optical only recursive Star-ID algorithm.}
    \label{fig:total_flow}
\end{figure}

The proposed algorithm is a synergy between QuateRA and RSI. QuateRA's adaptive angular velocity estimation is used in the RSI to estimate the attitude, and the RSI's attitude estimate is used in QuateRA to estimate the angular velocity. However, in order to get an initial estimate of the angular velocity, at least two quaternions are needed. Therefore, the lost-in-space case must be solved twice before beginning the proposed algorithm. The overall technique is summarized in the flowchart displayed in Figure \ref{fig:total_flow}.

In this algorithm, first, the variable $k$ is initialized as zero and will denote the number of valid frames. Next, a ``new frame" (or a new picture) is taken that provides the observed stars in the camera frame, $B_{k+1}$, and the timestamp $t_{k+1}$. Next, the number of potential stars is checked to ensure there are more than 3 stars present (note, in this step we are not checking for spikes which are handled in the lost-in-space algorithm (LISA) and the recursive Star-ID (RSI) algorithm). If the number of potential stars is not greater than three the algorithm returns to the beginning and a new frame is taken, but $k$ is not incremented. On the contrary, if the number of potential stars is greater than three, the algorithm then checks the current $k$ value. Since at initialization it is assumed no estimate of the angular velocity, $\hat{\B{\omega}}$, exists, the LISA must be used until $k > 1$. Once this condition is met, the RSI algorithm can be used instead of the LISA. After running either algorithm, a check for successful identification is conducted; if either Star-ID technique fails, the algorithm returns to the beginning and a new frame is taken without incrementing $k$. If the Star-ID is successful, the $R_{k+1}$ inertial vectors associated with the $B_{k+1}$ observations are used in an attitude estimation technique to determine the quaternion $q_{k+1}$. This quaternion is then appended to the matrix $Q_k$, which contains previous quaternion estimations. Next, the value of $k$ is again checked to determine if it is greater than one. If this is not true, then there is only one quaternion measurement, and the algorithm returns to the beginning, increments $k = k + 1$, and takes another picture. If $k > 1$, the quaternion matrix is passed to the QuateRA algorithm, which estimates the angular velocity vector, $\hat{\B{\omega}}_{k+1}$. This value and the current time step's data (i.e. $B_{k+1}, I_{k+1}, \hat{\B{\omega}}_{k+1}$, and  $t_{k+1}$) are stored and passed to the RSI algorithm for the subsequent loop. Next, a stopping criteria is checked, this can either be based on the number of valid measurements, $k$, the maximum observation time, or other user-defined criteria. If this check is satisfied, the algorithm terminates, if not, the algorithm returns to the beginning, increments $k = k + 1$, and takes another picture. 

\subsection{Adaptive Window}

While the QuateRA algorithm was based on pure spin dynamics, the algorithm can also be used when the angular velocity vector is near pure spin for the relative sample frequency. In other words, if sampled at a fast enough frequency, neighboring quaternion measurements will lie nearly on the same plane and can be used for estimation. This process is enabled by defining a ``sliding window'' of quaternion measurements to be used in the QuateRA algorithm. The SVD is computed for the quaternion sequence described in Eq.  \eqref{eq_avast:meas_matrix}, where the third singular value, $\sigma_3$, can be used to define how accurately the quaternion plane describes the dynamics or, in other words, how accurate is the pure spin assumption for the set of quaternion considered. If $\sigma_3$ is greater than some tolerance $\varepsilon_{\sigma_3}$, then the window is reduced by one quaternion measurement (i.e. the value of $n$ in Eq.  \eqref{eq_avast:meas_matrix} is decreased by one), until either, $\sigma_3 < \varepsilon_{\sigma_3}$, or only two quaternion measurements remain. However, careful attention must be given to the limiting case of two quaternion measurements, since a reduction to only two measurements could imply that the measurement frequency needs to be increased. Case 4 of the numerical results uses the sliding window approach where $\varepsilon_{\sigma_3} = 1 \times 10^{-9}$ is selected as the tolerance.

\section{Numerical Validation}

In this section, we present four unique tests to validate the accuracy and speed of the proposed method. For all simulations, the camera parameters detailed in Section \ref{sec:camera} were used to generate the star field provided by the Hipparchus star catalogue. Next, in Section \ref{sec:compare}, a speed comparison of the recursive algorithm with respect to the Pyramid Star-ID technique \cite{Pyramid} is conducted as a performance benchmark. Following this, Sections \ref{sec:case1}, \ref{sec:case2}, \ref{sec:case3}, and \ref{sec:case4} highlight situations where the recursive algorithm along with QuateRA \cite{QuateRA} can be utilized in the absence of rotational sensors. For all cases, an adaptive window for the number of quaternions used in QuateRA was selected such that the number of measurements was maximized and that the two smallest singular values of the SVD ($\sigma_3$ and $\sigma_4$) were both less than $1 \times 10^{-9}$. Table \ref{tab:testSummary} summarizes the key differences between these four tests.

\begin{table}[H]
\centering
\caption{Summary of test features}
\begin{tabular}{ccc} 
\toprule
{} & {Spin Axis} & {Spin Rate} \\ \midrule
{Case 1}  & {Constant/Unknown} & {Constant/Known} \\
{Case 2}  & {Constant/Unknown} & {Constant/Unknown} \\
{Case 3}  & {Constant/Unknown} & {Time Varying/Unknown} \\
{Case 4}  & {Time Varying/Unknown} & {Time Varying/Unknown} \\
\bottomrule
\end{tabular}
\label{tab:testSummary}
\end{table}

\subsection{Camera parameters for numerical tests}\label{sec:camera}
Table \ref{tab:param} shows the camera parameters used for all of the following simulations,
\begin{table}[ht]
    \centering
    \caption{Star tracker parameters}\label{tab:param}
    \begin{tabular}{cc}
        \toprule
        Virtual Star Tracker Parameter & Value \\ \midrule
        $3\sigma$ centroid error & $10$ arcseconds \\
        Magnitude threshold & $5.0$ \\
        Spike probability & $U[0,5]$ \\
        Number of rows & $1,024$ \\
        Number of columns & $1,024$ \\
        Pixel pitch & $0.018$ mm \\
        Focal length & $50.47$ mm \\ \bottomrule
    \end{tabular}
\end{table}
where $U[0,5]$ represents the uniform distribution of integers in the range $[0,5]$.
Note that for these simulations, specific CCD images are not being generated; rather, the star field is being perturbed using the centroiding error specified in Table \ref{tab:param}. Additionally, the parameter ``spike probability'' relates to the number of spikes, $N_s$, randomly generated in an image.

\subsection{Speed comparison with Pyramid}\label{sec:compare}
The major benefit of the proposed RSI technique is the computational time. Assuming an accurate estimation of the angular velocity of the camera, the $k$ frame can be utilized to identify the stars in the $k+1$ frame according to the process summarized in Figure \ref{fig:flow}. If the number of recurrent stars is greater or equal to three, then these can be used as a reference triangle to identify all of the other stars in the frame and thus reduce the computational complexity (i.e. identifying recurrent stars is less computationally intensive than identifying the original ``Pyramid'' in the Pyramid algorithm). As a way to quantify the speed gained when using the recursive technique, the algorithm presented in this paper was compared to Pyramid \cite{Pyramid}. For all cases, the speed tests were performed in C++ on a MacBook Pro (2016) macOS Version 10.15.3, with a 3.3 GHz Dual-Core Intel\textsuperscript{\textregistered} Core\texttrademark \, i7 and with 16 GB of RAM. All run times were calculated using the system\_clock function in the C++ boost chrono library.

First, the algorithm was tested in the best case scenario, where all of the identified stars in frame $k$ are recurrent in frame $k+1$. This was implemented by setting the angular velocity to zero in the simulation, which effectively causes frames $k$ and $k+1$ to be identical, aside from the spikes. However, in order to also test robustness, the number of spikes per frame was varied from 0 to 10. The results of this test averaged over 100,000 runs are presented in Table \ref{tab:bestCaseRecursive}.

\begin{table}[ht]
\centering
\caption{Speed comparison between recursive Star-ID and Pyramid in the best case recursive scenario (all stars in new frame were identified in the previous frame) averaged over 100,000 runs.}
\begin{tabular}{cccc} 
\toprule
\makecell{Number \\ of Spikes} & \makecell{Pyramid \\Speed \cite{Pyramid} [$\mu s$]} & \makecell{Recursive \\ Speed [$\mu s$]} & {Pyramid/Recursive}\\ \midrule
{0}  & {79.5} & {3.16} & {25.2}\\
{1}  & {87.5} & {4.81} & {18.2}\\
{2}  & {97.2} & {6.32} & {15.4}\\
{3}  & {103.3} & {7.49} & {13.8}\\
{4}  & {112.9} & {8.92} & {12.7}\\
{5}  & {122.0} & {10.2} & {11.9}\\
{6}  & {129.8} & {11.6} & {11.2}\\
{7}  & {137.8} & {12.9} & {10.7}\\
{8}  & {147.7} & {14.2} & {10.4}\\
{9}  & {156.9} & {15.6} & {10.1}\\
{10} & {166.8} & {16.9} & {9.89}\\
\bottomrule
\end{tabular}
\label{tab:bestCaseRecursive}
\end{table}

Analyzing these results, it can seen that the recursive algorithm outperforms Pyramid by an order of magnitude in computation time ranging from a speed gain (i.e. Pyramid/Recursive) of 25.2 to 9.89 times faster. Moreover, these results show that the recursive algorithm is robust to pixel spikes.

In addition to the best case test, the algorithm was also tested in the worst case scenario, which is synonymous with a poor estimation of the angular velocity. For this test, the recursive algorithm runs completely, fails, since three recurrent stars cannot be identified, and resorts to the lost-in-space algorithm, Pyramid, to identify the stars. The results, presented in Table \ref{tab:worstCaseRecursive} (averaged over 100,000 runs), depict the computational overhead when using the recursive algorithm, because the same Pyramid algorithm is called in both cases (e.g. after the recursive loop is run and fails). 
\begin{table}[ht]
\centering
\caption{Speed comparison between recursive Star-ID and Pyramid in the worst case recursive scenario (no recurrent stars are identified) averaged over 100,000 runs. In this case, the recursive algorithm is run, and when it fails, Pyramid is called.}
\begin{tabular}{cccc} 
\toprule
\makecell{Number \\ of Spikes} & \makecell{Pyramid \\Speed \cite{Pyramid} [$\mu s$]} & \makecell{Recursive \\ Speed [$\mu s$]} & {Pyramid/Recursive}\\ \midrule
{0}  & {77.3} & {78.7} & {0.98}\\
{1}  & {88.2} & {89.8} & {0.98}\\
{2}  & {96.4} & {98.1} & {0.98}\\
{3}  & {103.4} & {105.2} & {0.98}\\
{4}  & {113.1} & {115.2} & {0.98}\\
{5}  & {121.3} & {123.6} & {0.98}\\
{6}  & {128.2} & {130.5} & {0.98}\\
{7}  & {134.6} & {137.1} & {0.98}\\
{8}  & {142.5} & {145.1} & {0.98}\\
{9}  & {156.6} & {159.0} & {0.98}\\
{10} & {165.1} & {168.1} & {0.98}\\
\bottomrule
\end{tabular}
\label{tab:worstCaseRecursive}
\end{table}
Table \ref{tab:worstCaseRecursive} shows that this overhead is almost constant with respect to the number of spikes in the frame. In fact, for all test cases, the ratio of Pyramid/Recursive is approximately the same; when rounded to two decimal places, it is equal to 0.98 in all cases.

\subsection{Case 1: Fixed rover observation, ``Stellar Compass'' scenario}\label{sec:case1}
For this example consider the Stellar Compass scenario, where a system (e.g. rover, ascent vehicle, etc.) is equipped with a camera, two orthogonal inclinometers, and a precise clock. The purpose of this system is to estimate the geographical location in a GPS-denied environment. In this case, a position estimate can be obtained through a modern approach to the maritime technique, which utilizes a sextant. Furthermore, this approach is not limited to Earth, and becomes more important on the surface of other celestial bodies where there is not an established GPS. In general, the inclinometers provide the gravity direction (that is, the local horizon), the camera performs the Star-ID, and the clock identifies the body's orientation in the inertial frame. Therefore, since the system is attached and rotating with the body, it is also in pure spin about the bodies axis of rotation. In most cases (e.g., the Earth, other planets, and the moon), the angular speed is known with high accuracy and can be leveraged in the proposed algorithm. The following test simulates the Stellar Compass scenario for a system on Earth assuming a known angular speed and where the angular velocity vector is constant, but unknown. 

\begin{figure}[ht]
    \centering\includegraphics[width=.8\linewidth]{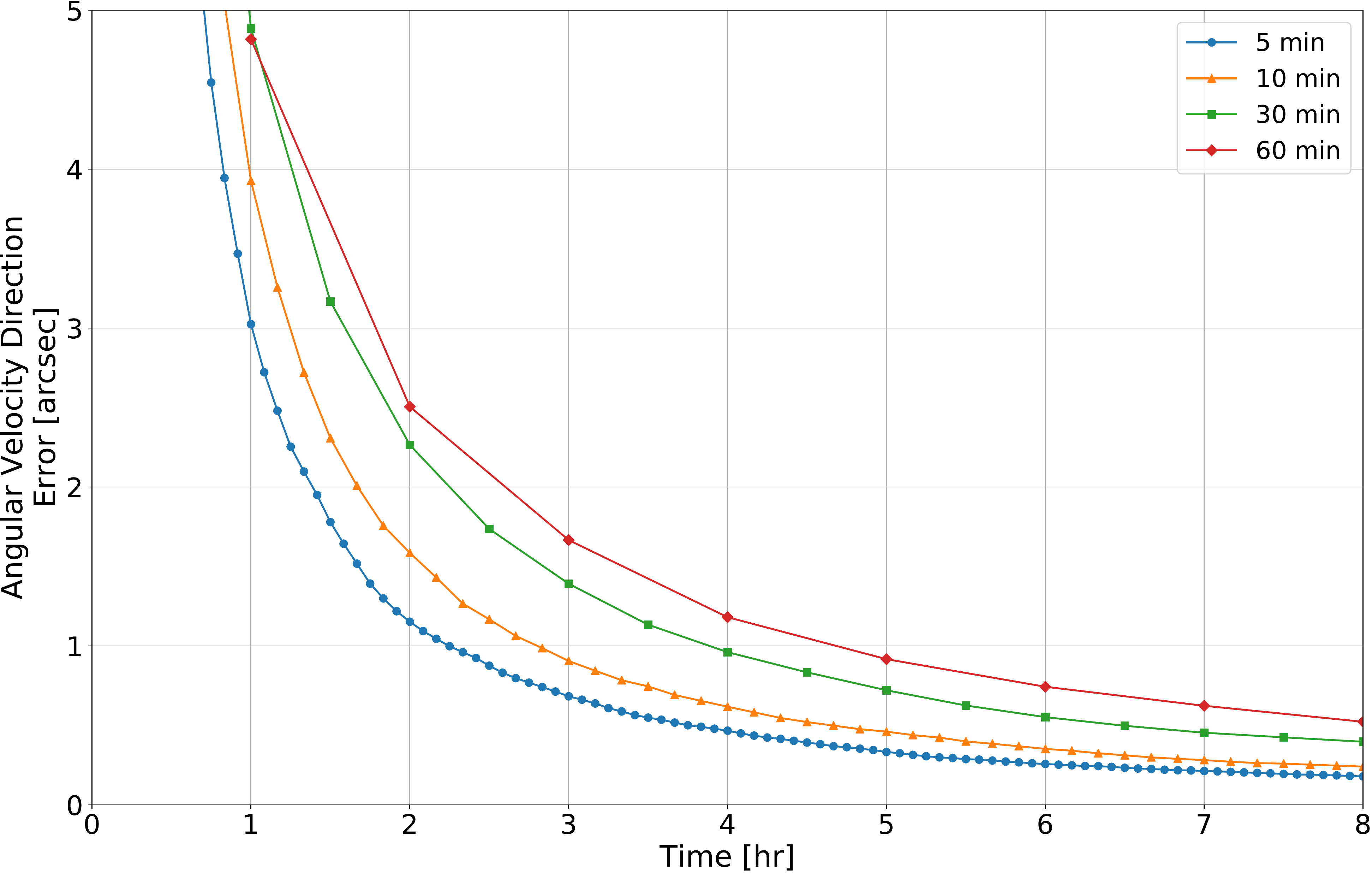}
    \caption{Results of the 1,000 trial Monte Carlo simulation for the `Stellar Compass'' scenario. The axis of rotation (given in arcseconds) is plotted over the duration of observation. Additionally, the results are split into four measurement frequencies quantified as 5 (circle), 10 (triangle), 30 (square), 60 (diamond) minutes between measurements.}
    \label{fig:spsAngErr}
\end{figure}

To test this scenario, a Monte Carlo simulation of 1,000 trials was constructed where the position was randomly initialized on the surface of the planet with the camera pointing in the zenith direction. An observational period of 8 hours was considered over a swath of 4 different measurement frequencies (5, 10, 30, and 60 minutes between measurements). The results of this test are presented in Figure \ref{fig:spsAngErr} and Table \ref{tab:spsPyramidCall}. In Figure \ref{fig:spsAngErr}, the error of the estimated axis of rotation is shown to be less than 5 arcseconds after one hour of measurements for all measurement frequencies. Additionally, after 5 hours, the error reduces below 1 arcsecond. 

As a way to determine the effectiveness of the recursive Star-ID technique, the percentage of Pyramid calls was determined and is displayed in Table \ref{tab:geoPyramidCall}. It can be seen that for all measurement cases Pyramid is called less than 2\% of the time over the 1,000 Monte Carlo trials. 

\begin{table}[ht]
\centering
\caption{Monte Carlo test results of the ``Stellar Compass'' test case detailing how many times the recursive algorithm ``failed," which resulted in a call to the Pyramid algorithm. Of the 1,000 test cases per picture frequency Pyramid was always called < 2\% of the time.}
\begin{tabular}{cc} 
\toprule
\makecell{Measurement Frequency\\ $[$min/measurement$]$ } & \makecell{Percent of \\ Pyramid Calls $[$\%$]$}\\ \midrule
{5}   & {1.1}\\
{10}  & {0.4}\\
{30}  & {0.1}\\
{60}  & {0.8}\\
\bottomrule
\end{tabular}
\label{tab:spsPyramidCall}
\end{table}

\subsection{Case 2: Satellite in geosynchronous orbit}\label{sec:case2}
\begin{figure}[ht]
    \centering\includegraphics[width=.8\linewidth]{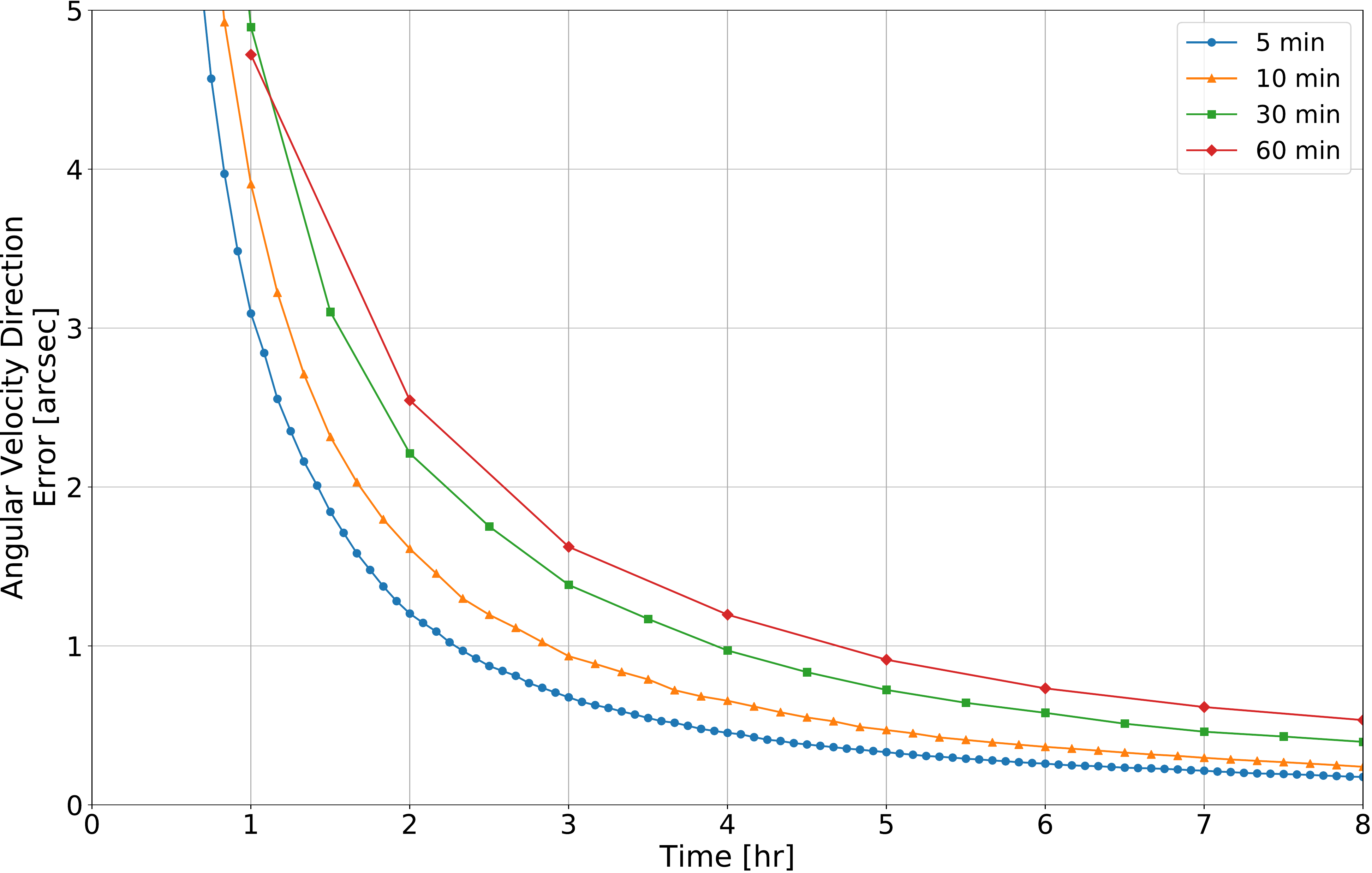}
    \caption{Results of the 1,000 trial Monte Carlo simulation for the geosynchronous satellite scenario. The axis of rotation (given in arcseconds) is plotted over the duration of observation. Additionally, the results are split into four measurement frequencies quantified as 5 (circle), 10 (triangle), 30 (square), 60 (diamond) minutes between measurements.}
    \label{fig:geoAngErr}
\end{figure}

The following example considers a satellite in geosynchronous orbit where both the angular speed and direction of the satellite are unknown. In this test, the recursive Star-ID success is now also highly dependent on an accurate estimate of the angular speed of the satellite during run-time. The accuracy of the axis of rotation estimate over this test is displayed in Figure \ref{fig:geoAngErr}, which is almost identical to the ``Stellar Compass'' scenario. This is to be expected, since the estimate of the axis of rotation is independent of the estimate of angular speed in the QuateRA algorithm.

\begin{figure}[ht]
    \centering\includegraphics[width=.8\linewidth]{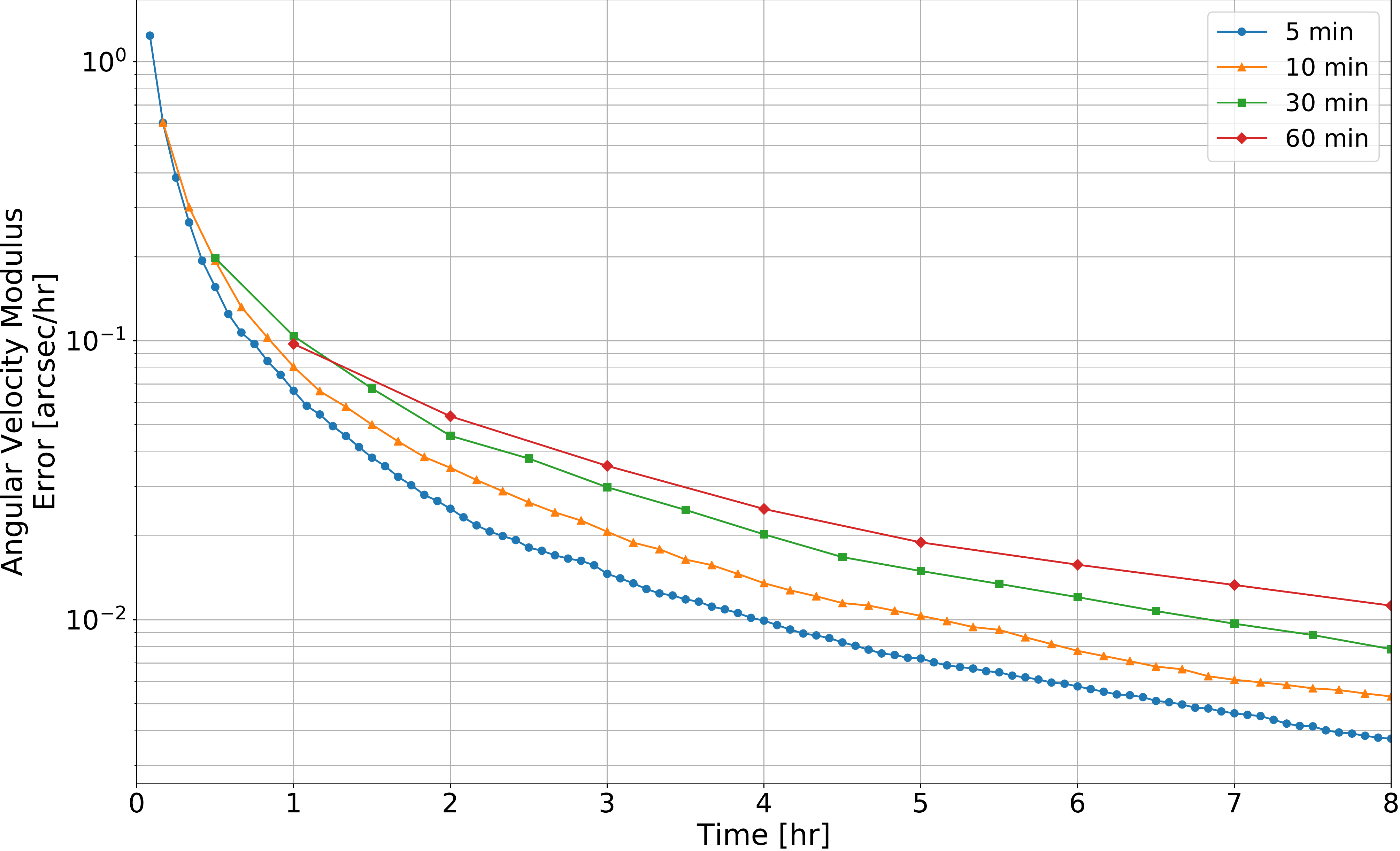}
    \caption{Results of the 1,000 trial Monte Carlo simulation for the geosynchronous satellite scenario. The angular speed (given in arcsecond/hour) is plotted over the duration of observation. Additionally, the results are split into four measurement frequencies quantified as 5 (circle), 10 (triangle), 30 (square), 60 (diamond) minutes between measurements.}
    \label{fig:geoModErr}
\end{figure}

Additionally, the angular speed (presented in arcseconds/hour) is displayed in Figure \ref{fig:geoModErr}. It can be seen that quickly after the initialization of the test, the error on this magnitude becomes insignificant, and therefore, not a factor in the success rate of the recursive Star-ID algorithm, which is more dependent in this tests on the estimate of the direction of rotation.

\begin{table}[ht]
\centering
\caption{Monte Carlo test results of the geosynchronous satellite detailing how many times the recursive algorithm ``failed," which resulted in a call to the Pyramid algorithm. Of the 1,000 test cases per picture frequency Pyramid was always called < 1\% of the time.}
\begin{tabular}{cc} 
\toprule
\makecell{Measurement Frequency\\ $[$min/measurement$]$ } & \makecell{Percent of \\ Pyramid Calls $[$\%$]$}\\ \midrule
{5}   & {0.4}\\
{10}  & {1.0}\\
{30}  & {0.5}\\
{60}  & {0.3}\\
\bottomrule
\end{tabular}
\label{tab:geoPyramidCall}
\end{table}

\subsection{Case 3: Bang-bang satellite reorientation}\label{sec:case3}
Since the QuateRA algorithm is valid for any planar rotation (since it is an SVD filter determining the quaternion plane of rotation), it can be used in simple spacecraft maneuvers, which includes single axis spacecraft reorientation maneuvers. For this test, consider a bang-bang control where the rotation angle $\theta$ is a function of time,
\begin{equation}\label{eq:control}
    \theta(t) = \begin{cases} \frac{1}{2}u t^2 \quad &\text{for} \quad t \leq \frac{t_f}{2} \\ -\frac{1}{4} \left(2 t^2 - 4 t t_f + t_f^2 \right) u \quad &\text{for} \quad t > \frac{t_f}{2} \end{cases}
\end{equation}
where the equations specify $\theta(0) = \dot{\theta}(0) = \dot{\theta}(t_f) = 0$, $t_f$ is the total time of the maneuver, and $u$ is the control. In order to determine the values of the final time and control we apply some restrictions on the maneuver. Let us assume that the desired control must meet the following conditions, $\theta(t_f) = \theta_f$ and $\dot{\theta} \leq \dot{\theta}_{\max}$ Therefore, using the function defined in Eq.  \eqref{eq:control}, the control is found to be,
\begin{equation*}
    u = \frac{4 \theta_f}{t_f^2}
\end{equation*}
and the final time is specified as
\begin{equation*}
    t_f = \frac{2\theta_f}{\dot{\theta}_{\max}}.
\end{equation*}
Therefore, by specifying the final orientation and the maximum allowable rotation rate, the piecewise function given in Eq.  \eqref{eq:control} gives us the analytical expression for orientation and angular speed, which can be used to construct the simulated maneuver for the recursive Star-ID algorithm.

\begin{figure}[H]
    \centering\includegraphics[width=0.8\linewidth]{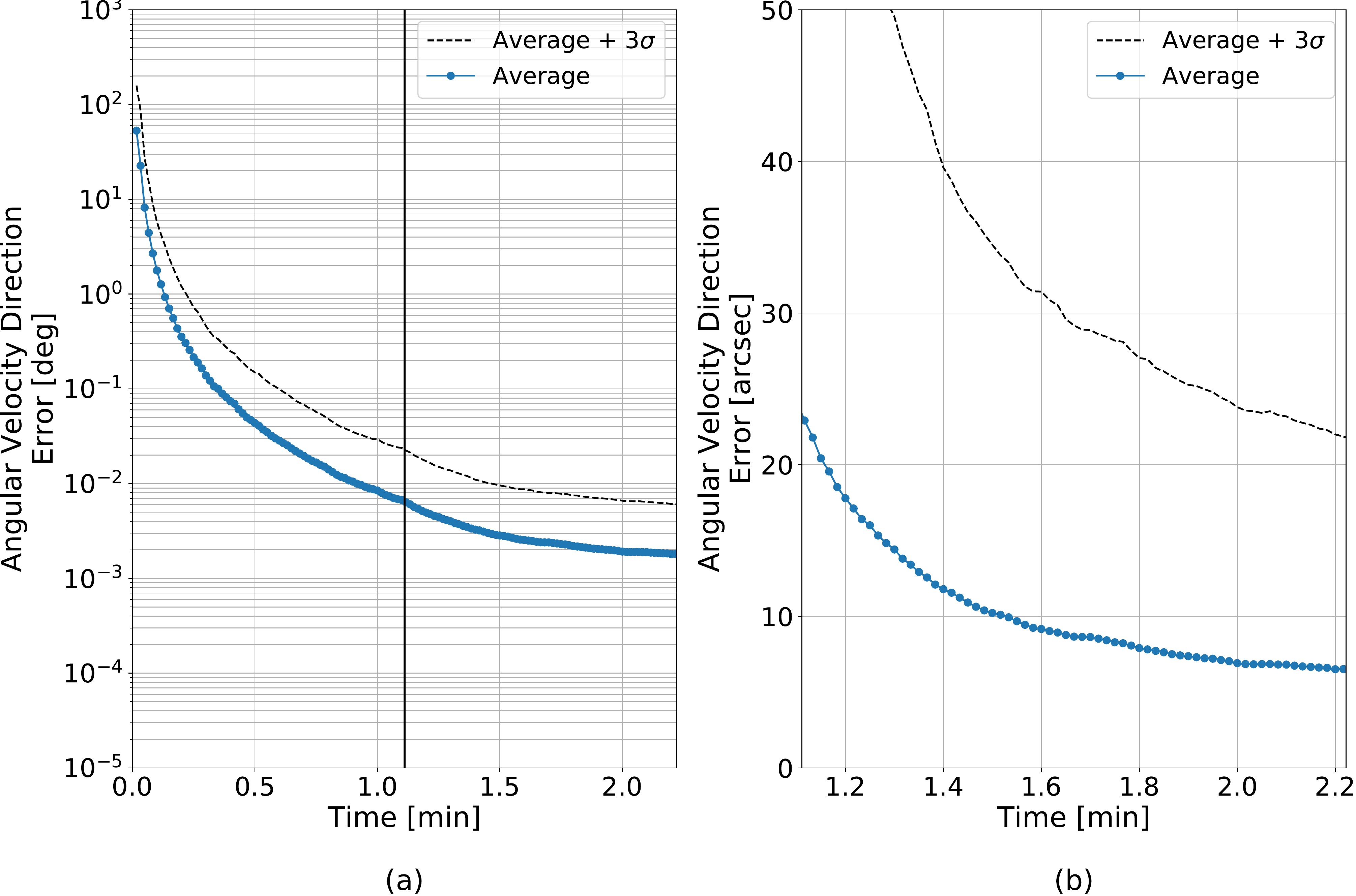}
    \caption{Results of the 1,000 trial Monte Carlo simulation for the bang-bang single axis spacecraft reorientation. The results shown in (a) show the angular velocity direction error (given in degrees) over the entire maneuver. The dashed line signifies $3\sigma$ limit of the average, and the vertical line represents the exact time the control switches. Additionally, (b) is a zoomed in view of the same test displaying only the second half of the maneuver. Notice that the scale of this plot is in arcseconds.}
    \label{fig:singleAxisAngErr}
\end{figure}

\begin{figure}[ht]
    \centering\includegraphics[width=0.8\linewidth]{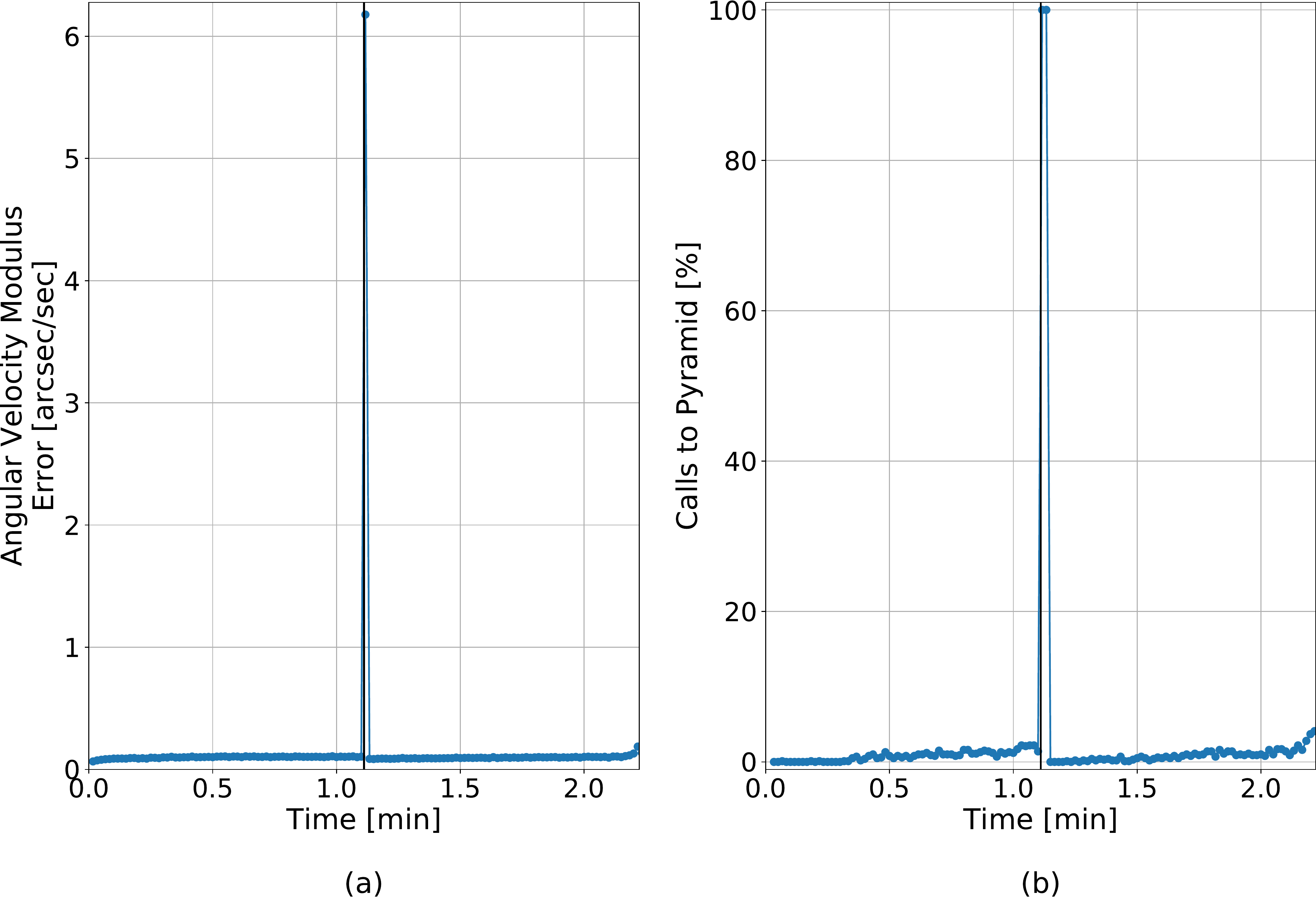}
    \caption{Results of the 1,000 trial Monte Carlo simulation for the bang-bang single axis spacecraft reorientation displaying (a) the angular velocity modulus error and (b) the average number of calls to Pyramid.}
    \label{fig:singleAxisModErr_Pyramid}
\end{figure}

For this specific test, the values of $\theta_f = 10$ [deg] and $\dot{\theta}_{\max} = 0.15$ [deg/sec] were selected as the control parameters and a Monte Carlo simulation of 1,000 trials was conducted where the attitude was randomly initialized. Additionally, during these maneuvers, the measurements were taken at a frequency of 1 Hz. The results of this test are presented in Figures \ref{fig:singleAxisAngErr} and \ref{fig:singleAxisModErr_Pyramid}, which detail the accuracy and robustness of the proposed algorithm. In Figure \ref{fig:singleAxisAngErr}(a) the angular velocity direction error is presented over the maneuver time where the vertical line signifies the switch of the control.
In order to better interpret the accuracy of the method at the end of the maneuver, a zoomed in and rescaled (the units have been change to arcseconds), plot is provided in Figure \ref{fig:singleAxisAngErr}(b). Here it is clear that the axis of rotation, on average, is estimated to be < 10 arcseconds at the end of the maneuver with a plus $3\sigma$ value slightly larger than 20 arcseconds. The reduction in the accuracy compared to the other tests is due to the reduced amount of total rotation. In this test, the maneuver only spans a range of $10$ degrees. This causes the quaternions in the plane to be closer together, reducing the accuracy of the SVD filter. 

The error in angular speed is presented in Figure \ref{fig:singleAxisModErr_Pyramid}(a), where all estimates are less than 1 arcsecond/sec, except the estimate immediately following the switch in control. This behavior is to be expected because there is a sharp change in the speed at this point in the maneuver. This result can also be observed in the plot detailing the percentage of times Pyramid was called during the maneuver shown in Figure \ref{fig:singleAxisModErr_Pyramid}(b).
This plot shows that directly after the control switch, Pyramid is always called twice before the recursive Star-ID is able to resume self-sufficiency; it takes two measurements to reinitialize the recursive algorithm.

\subsection{Case 4: Time varying spacecraft dynamics}\label{sec:case4}
The final test was constructed to highlight the capabilities of the algorithm when conditions deviate from the pure spin case. Therefore, consider an angular velocity that is changing in both direction and magnitude given by the inertial angular velocity vector (in degrees/second) defined as,
\begin{equation*}
    \B{w} = \begin{cases}\begin{Bmatrix} 3, & 0, & 0 \end{Bmatrix}\T \quad &\text{if} \quad t < t_f/4 \\ \begin{Bmatrix} 3, & 0.003\left(t - \frac{t_f}{4}\right), & -0.0015\left(t - \frac{t_f}{4}\right)\end{Bmatrix}\T &\text{if} \quad t \geq t_f/4 \end{cases}
\end{equation*}
where $t_f$ is the simulation final time, which in the numerical test is defined as 2 minutes. In this case, the spacecraft is in pure spin, which linearly changes after a fourth of the final time. For the recursive algorithm, since the dynamics are not pure spin, the sliding window was used for the quaternion measurements with a tolerance of $\varepsilon_{\sigma_3} = 1 \times 10^{-9}$.

\begin{figure}[!htbp]
    \centering\includegraphics[width=0.9\linewidth]{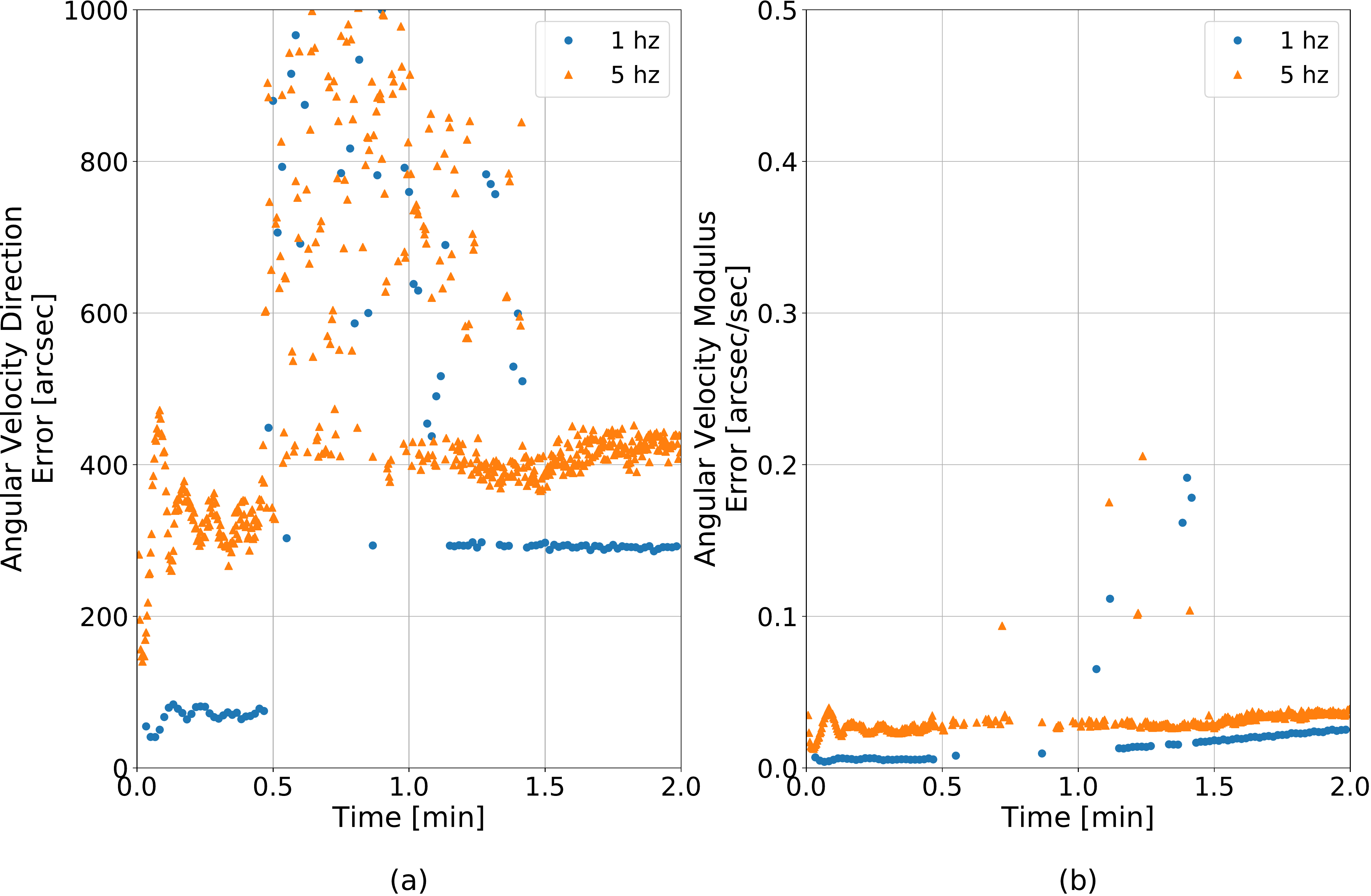}
    \caption{Results of the 1,000 trial Monte Carlo simulation for the varying spacecraft dynamics test displaying (a) the angular velocity direction error and (b) the angular velocity modulus error.}
    \label{fig:adaptiveAngModErr}
\end{figure}

\begin{figure}[!htbp]
    \centering\includegraphics[width=0.9\linewidth]{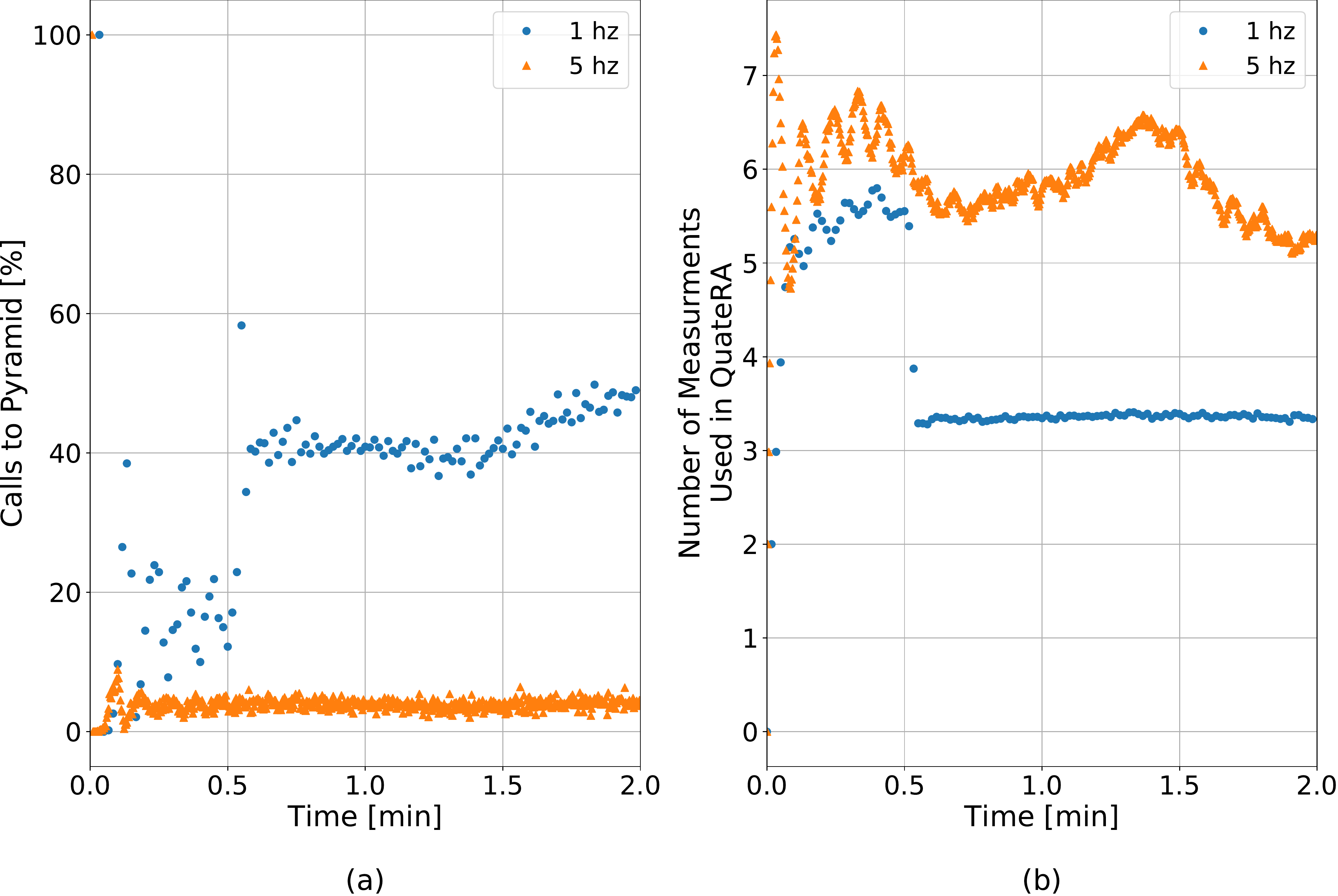}
    \caption{Results of the 1,000 trial Monte Carlo simulation for the varying spacecraft dynamics test displaying (a) the average number of call to Pyramid  and (b) average number of quaternions used in QuateRA with the tolerance of $\varepsilon_{\sigma_3} = 1 \times 10^{-9}$  .}
    \label{fig:adaptivePyramidQuateRA}
\end{figure}

For this case, a Monte Carlo simulation of 1,000 trials was constructed with the same parameters as the prior tests. First, the error in the direction is reported in Figure \ref{fig:adaptiveAngModErr}(a) and the corresponding error in magnitude is given in \ref{fig:adaptiveAngModErr}(b). By quick inspection, it can be seen that sampling at a lower frequency produces a slightly better result in both direction and magnitude error. This result is obtained because at higher frequencies the attitude (and therefore the quaternion) is not varying as much causing the measurements to lie closer in the plane. However, the results of the 1 Hz test are worse when the dynamics change, when the simulation reaches 0.5 minutes. This can be seen in Figure \ref{fig:adaptiveAngModErr}(b) where there is a gap in the 1 Hz data where the error is greater than 0.5 arcseconds per second (the method loses one to two orders of magnitude in accuracy during this time).
However, the difference in frequency has significant results on the number of times Pyramid is called and the number of quaternions used in each QuateRA call over the test. This is displayed in Figure \ref{fig:adaptivePyramidQuateRA}. In Figure \ref{fig:adaptivePyramidQuateRA}(a) it can be seen that at 1 Hz, the algorithm is still highly reliant on Pyramid where in the dynamic portion of the simulation the proposed algorithm is observed to rely on Pyramid roughly 40\% of the time. In contrast, sampling at 5 Hz reduces this to about 5\%. In Figure \ref{fig:adaptivePyramidQuateRA}(b) it can be seen that during the dynamic portion of the simulation the 1 Hz case relies solely on approximately three quaternion measurements. This is a sign that the dynamics change too fast for that sampling frequency. On the other hand, for the 5 Hz, case it can be seen that the average number of quaternions used is around six quaternion measurements.

\section{Conclusions}

In this paper a new recursive Star-ID algorithm is presented and validated over four test scenarios. This recursive algorithm takes advantage of the accuracy and adaptability of the recent ``Quaternion Regression Algorithm,'' to estimate the angular velocity. This recursive algorithm is robust to the presence of spikes (fake stars), such as planets, reflecting satellites or debris, electronic noise, or simply stars not included in the on-board star catalog. The accurate angular velocity estimation allows for a precise prediction of the stars' centroids, and establishes the basis to recursively identify stars in subsequent frames.

The algorithm was compared with Pyramid (the current state-of-the-art lost-in-space algorithm), where its speed gain ranged from 10 to 25 times faster in the best case scenario to a 2\% speed reduction in the worst case. Additionally, four unique simulations prove that this algorithm can be used in a variety of scenarios with different spacecraft dynamics.

\vspace{6pt} 

 
\appendix
\section{Pyramid Star Identification}\label{app:Pyramid}
The Pyramid star identification algorithm is briefly summarized in the three steps that follow. Note that in these steps stars are referenced using an alphabetic character, such as the $i$, $j$, and $k$ stars, while star triangles are referenced using curly braces, such as the $\{i,j,k\}$ star triangle. For a more detailed explanation of the algorithm consult Ref. \cite{Pyramid}.
\begin{enumerate}
    \item First, Pyramid searches for a unique star triangle. A unique star triangle is any triangle composed of three stars whose interstellar angles could only form that particular triangle. The triangle is identified by selecting all star pairs that match, within an assigned tolerance, the three observed star pairs of the triangle. The $k$-vector \cite{original, Neta} is used to perform these range searches quickly, and is the engine that enables Pyramid to perform the Star-ID in real time. Let the unique star triangle be composed of three stars; call them the $i$, $j$, and $k$ stars. If no unique star triangles are found, then Pyramid reports it cannot identify any stars. 
	\item Next, Pyramid searches for a reference star, call it star $r$. A reference star $r$ is any star in the field-of-view such that the $\{i,j,r\}$, $\{i,k,r\}$, and $\{j,k,r\}$ star triangles are unique star triangles. If a reference star is found, then the $i$, $j$, $k$, and $r$ stars are considered identified. These four stars represent vertices of a ``Pyramid,'' hence the name of the algorithm. If a reference star cannot be found, the algorithm returns to step one and tries to find a different unique star triangle. 
	\item Finally, the remaining stars in the field-of-view are identified using the same process as the reference star $r$. For a given star $s$, if the $\{i,j,s\}$, $\{i,k,s\}$, and $\{j,k,s\}$ star triangles are unique star triangles, then star $s$ is identified. Otherwise, star $s$ is discarded. 
\end{enumerate}

\bibliographystyle{unsrt} 
\bibliography{mybib} 
\end{document}